\documentclass[twocolumn]{aastex62}

\usepackage{subfigure}
\usepackage{amsmath}

\begin{document}


\title{Radial velocity and chemical composition of evolved stars in the  open clusters NGC~6940 and Tombaugh~5}



\correspondingauthor{Martina Baratella}
\email{martina.baratella.1@studenti.unipd.it}

\author{Martina Baratella}
\affiliation{Dipartimento di Fisica e Astronomia {\it Galileo Galilei}, 
Vicolo Osservatorio 3,
I-35122, Padova, Italy}

\author{Giovanni Carraro}
\affiliation{Dipartimento di Fisica e Astronomia {\it Galileo Galilei}, 
Vicolo Osservatorio 3,
I-35122, Padova, Italy}
\affiliation{Special Astrophysical Observatory, Russian Academy of Sciences,
 Nizhny Arkhyz 369167, Russia}

\author{Valentina D'Orazi}
\affiliation{INAF, Padova Astronomical Observatory, Vicolo Osservatorio 5,
 I-35122, Padova, Italy}

\author{Eugene A. Semenko}
\affiliation{Special Astrophysical Observatory, Russian Academy of Sciences,
 Nizhny Arkhyz 369167, Russia}

\begin{abstract}
We present and discuss medium resolution (R $\sim$ 13000), high signal-to-noise ($\mathrm{\frac{S}{N}} \sim 100$), spectroscopic observations in the field of the open clusters NGC\,6940 and Tombaugh\,5. Spectra were recorded for seven candidate red giant stars in both clusters. For the latter we present the very first chemical abundance analysis. We derive radial velocities for all the stars in NGC\,6940, confirming membership to the cluster for all of them,  while on the same ground we exclude two stars in To\,5. We perform a chemical abundance analysis of different atomic species, in particular FeI, SiI, CaI, TiI and NiI. The mean metallicity of NGC\,6940 is [Fe/H]=+0.09$\pm$0.06\,dex, in good agreement with previous works, while for To\,5 is  [Fe/H]=+0.06$\pm$0.11\,dex. Therefore, both clusters exhibit a chemical composition close to the solar value, and do not deviate from the [Fe/H] Galactic radial abundance gradient.
With these new values we estimate the fundamental cluster parameters, after having derived clusters' distances from the \textit{Gaia} DR2 database. By adopting these distances, we
derive updated estimated for the clusters ages:  1.0$\pm$0.1\,Gyr of NGC\,6940 and 0.25$\pm$0.05 Gyr for Tombaugh\,5. 
\end{abstract}


\section{Introduction} 
This paper continues a series started in Carraro et al. \citeyear{2016Car} that aims at expanding  the current sample of Galactic disk  intermediate age and old open clusters for which good metallicity and abundance ratios measurements are available. As for the Open Cluster Chemical Abundances from Spanish Observatories survey (OCCASO) (Casamiquela et al. \citeyear{2016C}) we are targeting clusters in the northern hemisphere, which are out of reach of the Gaia-ESO Spectroscopic Public Survey (Gilmore et al. \citeyear{2012G}).   In Carraro et al. \citeyear{2016Car} we presented the first abundance study of NGC 7762, a nearby old open cluster. Here we focus on two open clusters which are located at larger distance from the Sun, toward the anti-centre direction: NGC\,6940 and Tombaugh\,5 (hereafter To\,5).\\
We aim at finding and characterising open clusters in the Galacto-centric distance range (10-12 kpc) at which the Galactic radial abundance gradient has an abrupt change of slope (Twarog et al 1997, Sales Silva et al. 2015), and keeps flat outwards. In this way, we can expand the sample of fully characterised open clusters observable in the Northern hemisphere and improve the observational constraints of chemical evolution models.\\ 

\begin{figure}
\includegraphics[width=\columnwidth]{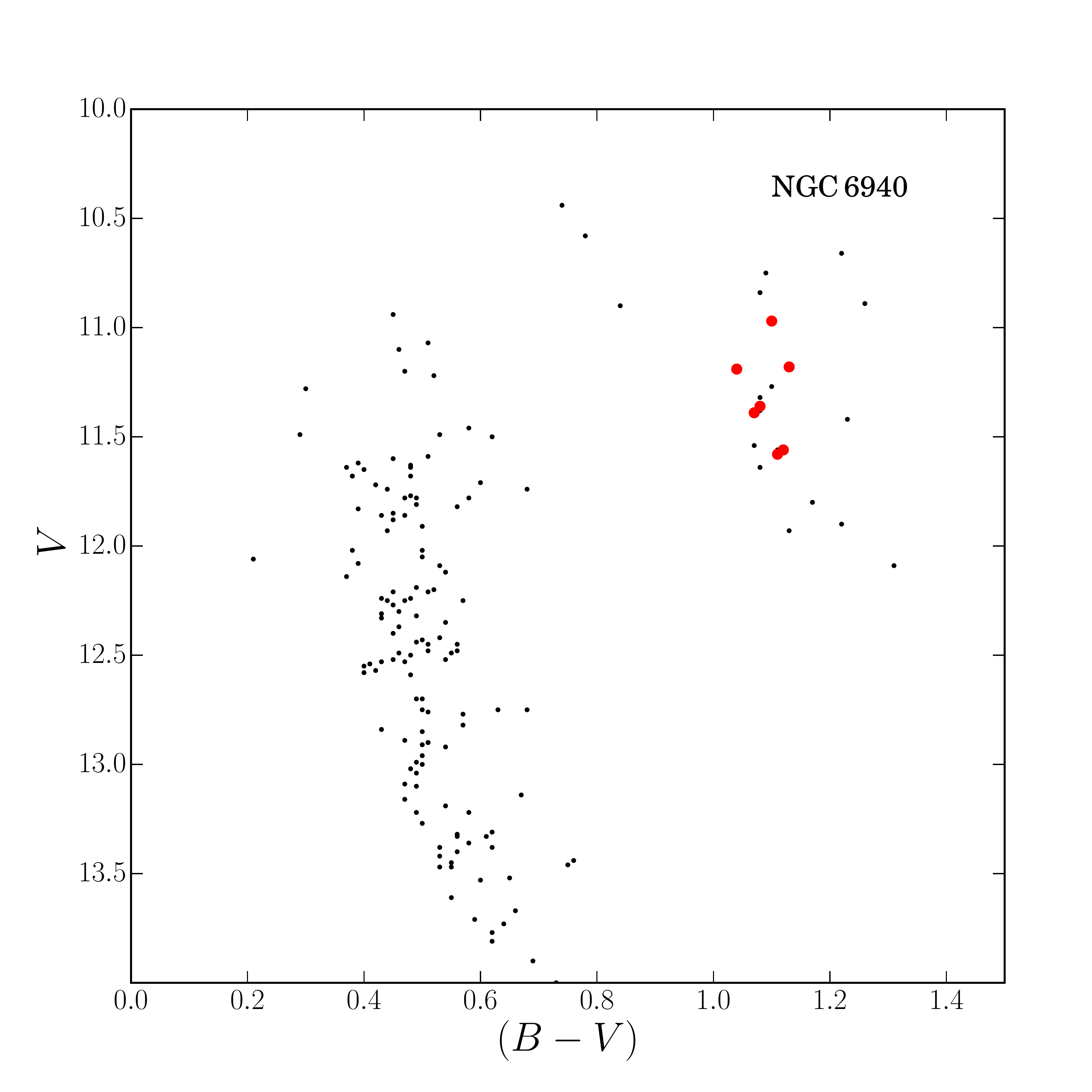}\quad\includegraphics[width=\columnwidth]{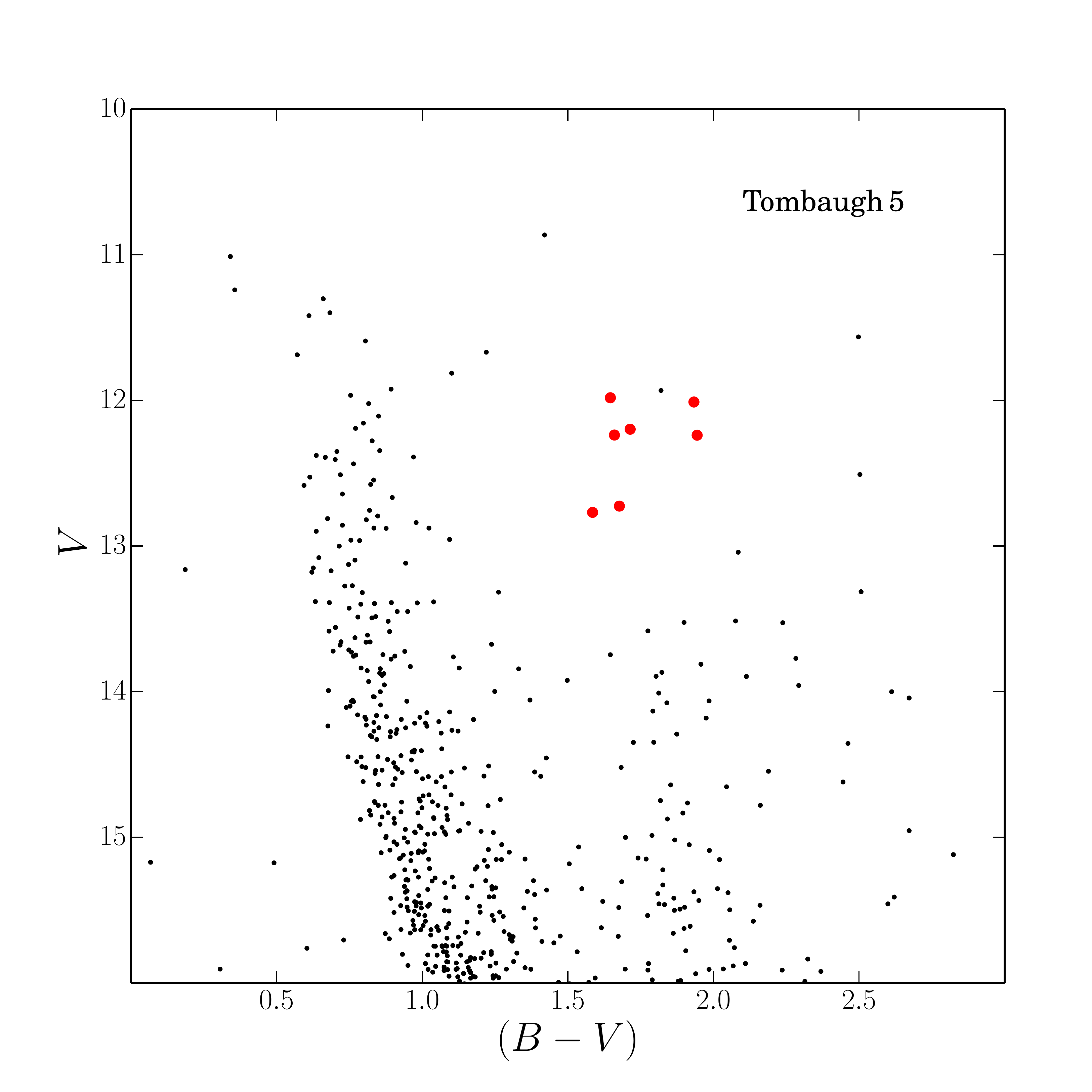}
\caption{Color magnitude diagrams of NGC\,6940 (Larsson-Leander (\citeyear{1960lar})) and To\,5 (Maciejewski \& Niedzielski (\citeyear{2007mac})). Red symbols indicate the stars for which we obtained spectroscopic data.}
\label{Fig1}
\end{figure}

\noindent
In this work, we present a spectroscopic investigation of the open clusters NGC\,6940 and To\,5. For the latter no radial velocity and chemical abundance analysis have been performed to date, while for the former abundances were available for a few stars only at the time of the observing proposal preparation.
 In addition, a literature scan indicates discrepant values
for To\,5 basic parameters (especially distance). 
We secured spectra for seven stars in both clusters, close to the expected location of red clump stars. We derived radial velocities and metallicities, as well as abundances ratios of $\alpha$ and iron-peak elements. The selected stars are shown in the colour magnitude diagrams (CMD) in Fig.\ref{Fig1} (highlighted in red) and their properties are summarised in Tab.\ref{Tab1}.\\

\noindent
The paper is organised as follows: in Sect. 2 we present a brief summary of  previous studies on NGC~6940 and To\,5, while in Sect.3 we present the observational material and the basic reduction strategy. Sect.4 is devoted to radial velocities measurements, while in Sect.5 we present the estimates of the stellar atmospheric parameters. The abundance analysis is presented in Sect.6 and Sect.7. Moreover,  in Sect.8 we derive updated estimates of the clusters' fundamental parameters. Finally, Sect.9 provides some discussion and summarises our findings.

\begin{table*}
\tabcolsep 0.3truecm
\caption{Basic information of the stars for which we obtained spectroscopic data. The signal-to-noise ratios ($\mathrm{\frac{S}{N}}$) in the last column have been measured using a few \AA~ continuum region at 6300\AA.}
\label{Tab1}
\centering
\begin{tabular}{lccccccccl}
\hline
\hline
ID &	RA(J2000.0)  &    Dec(J2000.0)	&  V	& B-V  &    J  &    H  &    K  &   $\mathrm{t_{exp}}$ (sec) &    $\mathrm{\frac{S}{N}}$ \\
\hline
\textbf{NGC\,6940} & & & & & & & & &\\
	28 & 20:33:25.0 & +28:00:46.9 &  11.56 & +1.12 & 9.565 & 9.049 & 8.929 & 2400 & 68 \\
	60 & 20:33:59.6 & +28:03:01.7 &  11.56 & +1.11 &  9.593 & 9.103 & 8.974 & 2400 & 92 \\
	67 & 20:34:04.1 & +28:16:48.6 &  11.18 & +1.13 & 9.134 & 8.608 & 8.518 & 2000 & 135\\
	108 & 20:34:25.6 & +28:13:41.5 &  11.19 & +1.04 & 9.275 & 8.857 & 8.696 & 2000 &110 \\
	130 & 20:34:38.7 & +28:20:22.7 &  11.39 & +1.07 & 9.453 & 8.966 & 8.853 & 2400 & 103 \\
	132 & 20:34:40.1 & +28:26:38.9 & 10.97 & +1.10 & 9.039 & 8.491 & 8.391 & 1800 & 125 \\
	139 & 20:34:47.6 & +28:14:47.3 &  11.38 & +1.08 & 9.464 & 8.942 & 8.804 & 2000 & 94 \\
\hline

	\textbf{To\,5} & & & & & &  & & & \\
	0006 & 03:48:17.46 & +59:11:19.68 &  12.198 & 1.714 & 8.888 & 8.222 & 7.960 & 2700 & 118\\
	5274 & 03:46:55.60 & +58:55:34.79 &  11.982 & 1.646 & 8.993 & 8.360 & 8.171 & 2700 & 139\\
	5521 & 03:48:47.41 & +59:02:28.18 &  12.238 & 1.660 & 9.231 & 8.483 & 8.279 & 2700 & 165\\
	7701 & 03:47:30.98 & +59:02:50.89 &  12.011 & 1.933 & 8.531 & 7.797 & 7.581 & 2700 &  111\\
	7834 & 03:47:49.92 & +58:56:20.84 &  12.769 & 1.585 & 9.624 & 9.015 & 8.782 & 3600 & 105\\
	8080 & 03:48:29.20 & +59:00:36.68 &  12.726 & 1.677 & 9.628 & 8.969 & 8.752 & 3600 &  86\\
	8099 & 03:48:32.98 & +59:15:16.63 &  12.239 & 1.944 & 8.651 & 7.936 & 7.699 & 2300 & 76\\

\hline\hline 
\end{tabular}
\end{table*}

\section{Overview of the literature on NGC~6940 and Tombaugh~5}

\subsection{NGC~6940}
NGC\,6940 is a rather well-known intermediate-age open cluster (see Tab.\ref{Tab2}).
The first membership assessment  was conducted by Vasilevskis \& Rach (\citeyear{1957vas}), who identified 216 possible members using proper motions. Later, Sanders (\citeyear{1972san}) confirmed the membership of 104 stars, implementing a new procedure based on the maximum likelihood method.\\
The first extensive photometric study was conducted by Larsson-Leander (\citeyear{1964lar}), who cross-correlated several previous studies and confirmed the estimates by Walker (\citeyear{1958wal}) and Johnson et al. (\citeyear{1961joh}) of the apparent distance modulus ((m-M)$_V$=9.5\,mag) and the reddening (E(B-V)=0.26, both of which
were later confirmed by Hartwick \& McClure (\citeyear{1972hart})). Jennens \& Helfer (\citeyear{1975jen}) derived the very first age estimate of the cluster assuming previous estimates of distance modulus and reddening, and finding an age of $\sim$500 Myr.\\
The first detailed spectroscopic study was conducted by Geyer \& Nelles (\citeyear{1985gey}) who derived radial velocities of 31 member stars, one of which is  included in our sample. Geisler (\citeyear{1988geis}) confirmed those estimates and determined a mean radial velocity of the cluster of 5.7$\pm$1.5\,km/s. Mermilliod \& Mayor (\citeyear{1989merm}) analyzed high-resolution spectra of member stars in different open clusters, among which NGC\,6940, confirming the existence in that region of at least six spectroscopic binaries, among which our \#130. Later, Mermilliod et al. (\citeyear{2008merm})  analysed 26 possible red giants stars, confirming 21 of them and deriving a mean cluster radial velocity of 7.89$\pm$0.14\,km/s.
We used all these works to compare our radial velocity measurements, as reported in Tab.\ref{Tab3}.\\
As for spectroscopic metallicity,  the first study was conducted by Thogersen et al. (\citeyear{1993thor}), who analyzed medium-resolution spectra of six red clump stars, confirming the previous estimates of the radial velocity and deriving a metallicity of -0.06$\pm$0.13\,dex. Among the most recent studies on metallicity of NGC\,6940, 
Blanco-Cuaresma et al. (\citeyear{2015bla}) and B$\mathrm{\ddot{o}}$cek-Topcu et al. (\citeyear{2016boc}) conducted a very detailed high-resolution spectroscopic study, obtaining values of the metallicity very similar, of respectively +0.04$\pm$0.09\,dex  (1 star) and +0.04$\pm$0.02\,dex (12 stars). Reddy et al. (\citeyear{2016red}) confirmed their measurements for three stars. We decided to observe NGC\,6940 since 
at the time of our proposal preparation only Reddy et al.(\citeyear{2016red}) and 
Blanco-Cuaresma et al. (\citeyear{2015bla}) studies were available.\\

\begin{table}
\tabcolsep 0.15truecm
\caption{Reddening, apparent distance modulus and typical red clump stellar masses used to derived the atmospheric parameters. The values are taken from WEBDA database.}
\label{Tab2}
\centering
\begin{tabular}{lccccc}
\hline\hline
Cluster   & l & b & E(B-V) & ${(m-M)_{\rm  V}}$ & m$_{\star}$ \\
&\footnotesize{($^{\circ}$)}& \footnotesize{($^{\circ}$)} & & & \footnotesize{(M$_{\odot}$)}\\
\hline
 NGC\,6940 & 69.860 & -7.147 & 0.21 & 10.10 & 2.05\\
 To\,5 & 143.942 & 3.573 & 0.80 & 13.70 & 3.50\\
\hline\hline
\end{tabular}
\end{table} 

\begin{figure*}
\centering
\subfigure[Spectra of NGC\,6940 stars.]{{\includegraphics[scale=0.25]{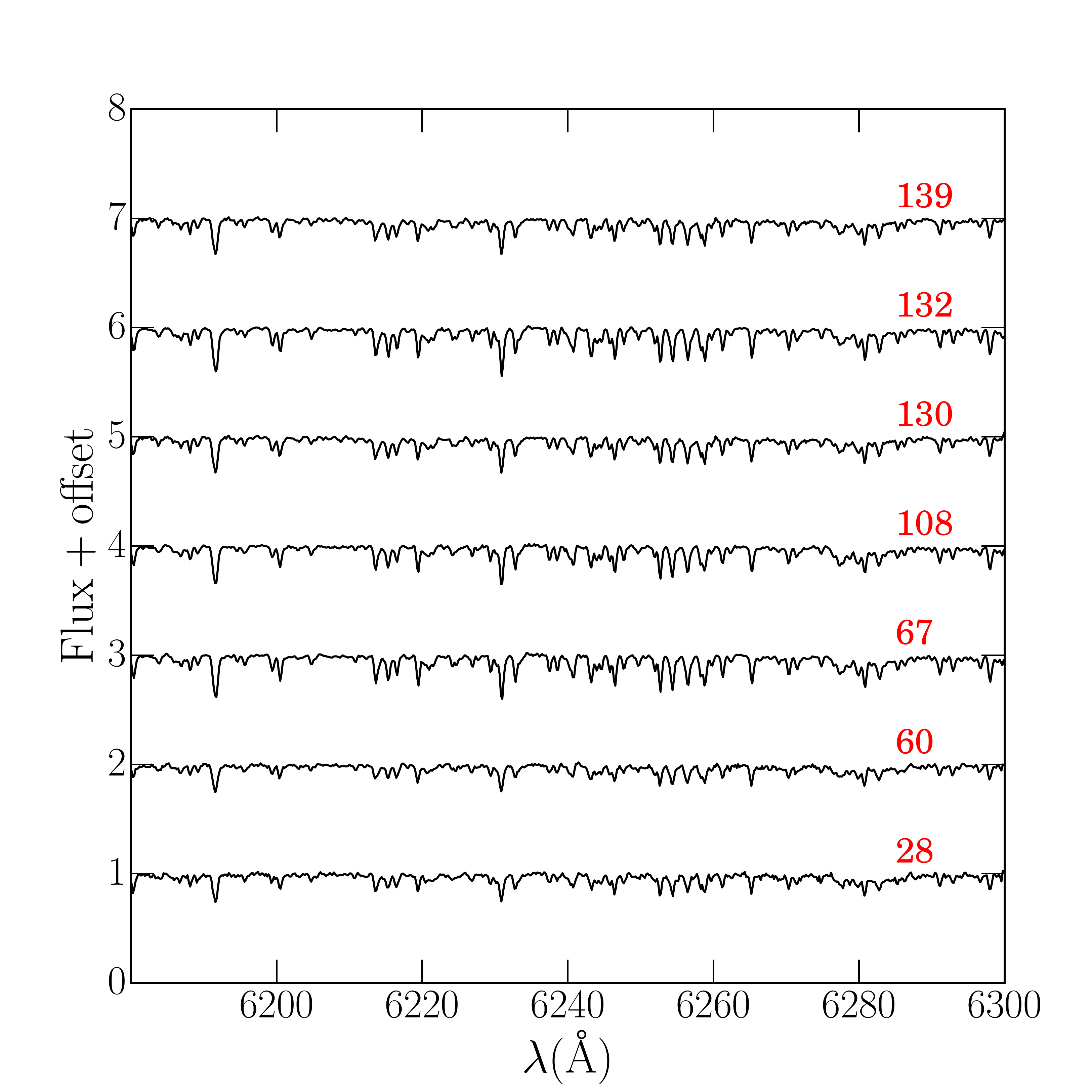}}}%
\qquad
\subfigure[Spectra of To\,5 stars.]{{\includegraphics[scale=0.25]{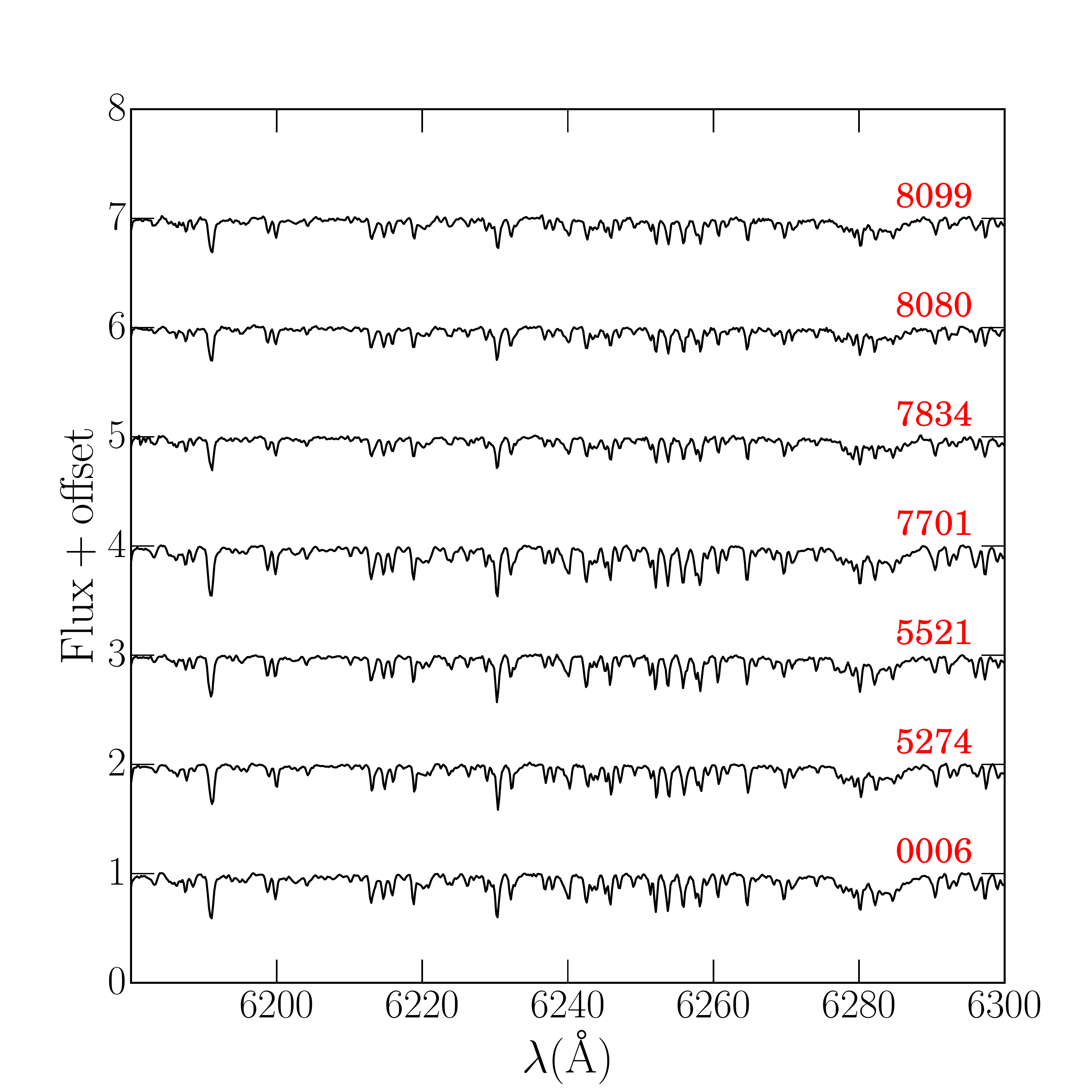}}}%
\caption{An excerpt of the individual spectra in the indicated wavelength range. Spectra are arbitrarily shifted in the vertical axis, for illustration purposes.}
\label{Fig2}
\end{figure*}

\subsection{Tombaugh\,5}

Unlike NGC\,6940, To\,5 is a poorly known star cluster, and its fundamental parameters are still unsettled.
It was discovered by Tombaugh (\citeyear{1941tom}), who provided the first estimate of the angular diameter of 17\,arcmin.\\
Only photometric analyses are available for it. 
There is a general agreement  on the cluster age ( $\sim$ 200-250\,Myr )
and color excess E(B-V) (0.80\,mag, Lata et al.(\citeyear{2004lat}), Maciejewski \& Niedzielski (\citeyear{2007mac}), Majaess, Turner \& Lane (\citeyear{2008maj}), Zdanavicius et al.(\citeyear{2011zda})). 
On the other hand, distance estimates crowd around two marginally discrepant values.
According to Lata et al.(\citeyear{2004lat}) BV study, the apparent distance modulus is around 13.7\,mag (which implies a heliocentric distance of 1.7\,kpc). This estimate was later confirmed by Zdanavicius et al.(\citeyear{2011zda}), who re-estimated the basic cluster parameters using photometric data in the Vilnius system.\\
Maciejewski \& Niedzielski (\citeyear{2007mac}) argue that Lata et al.(\citeyear{2004lat}) CMD was highly contaminated by field stars. By selecting the cluster putative members  in a statistical way, they obtained an apparent distance modulus of 13.10\,mag (which corresponds to a value of d$\sim$1.3\,kpc) using BV photometry as well. This value was later confirmed by Kharchenko et al. (\citeyear{2013kha}) via an 
automated fit of 2MASS photometry. 
A more careful analysis of the 2MASS CMD from  Majaess, Turner \& Lane (\citeyear{2008maj}) yields a distance of 1.7\,kpc.\\ 
No spectroscopic study is  available to our knowledge.  An estimate of the cluster metallicity, together with the recently released \textit{Gaia} DR2\footnote{This work has made use of data from the European Space Agency (ESA) mission
{\it Gaia} (\url{https://www.cosmos.esa.int/gaia}), processed by the {\it Gaia}
Data Processing and Analysis Consortium (DPAC,
\url{https://www.cosmos.esa.int/web/gaia/dpac/consortium}). Funding for the DPAC
has been provided by national institutions, in particular the institutions
participating in the {\it Gaia} Multilateral Agreement.} (Gaia Collaboration et al. \citeyear{2016gaia}; Gaia Collaboration et al. \citeyear{2018gaia}) astrometric and photometric data, should help to alleviate the aforementioned distance discrepancy.

\begin{table*}
\centering
\tabcolsep 0.55truecm
\caption{Radial velocity measurements for the program stars in NGC\,6940. Literature values are from Geyer \& Nelles (\citeyear{1985gey}, GN85), Geisler \citeyear{1988geis} (G88), Mermilliod \& Mayor \citeyear{1989merm} (M89), Thogersen et al. \citeyear{1993thor} (T93), Mermilliod et al. \citeyear{2008merm} (MMU08) and B$\mathrm{\ddot{o}}$cek Topcu et al. \citeyear{2016boc} (BT16).}
\label{Tab3}
\begin{tabular}{lccccccc}
\hline
\hline
ID &	$V_r$ & $V_{R,GN85}$ & $V_{R,G88}$ & $V_{R,M89}$ & $V_{R,T93}$ & $V_{R,MMU08}$ & $V_{R,BT16}$ \\
\hline
 & \footnotesize{(km/s)} & \footnotesize{(km/s)}&\footnotesize{(km/s)}&\footnotesize{(km/s)}&\footnotesize{(km/s)}&\footnotesize{(km/s)}&\footnotesize{(km/s)} \\
 \hline
28 & 6.47$\pm$1.90  & - & 3.2$\pm$3.2 & 8.12$\pm$0.28 & - & 7.99$\pm$0.16 & 8.90$\pm$0.08 \\
	60 & 8.41$\pm$1.30  &29$\pm$2.1 & - & 7.24$\pm$0.28 & - & 7.27$\pm$0.18 & 7.66$\pm$0.08\\
	67 & 8.38$\pm$0.47  & - & 10.6$\pm$3.5 & 7.67$\pm$0.21 & 7 & 7.69$\pm$0.19 & -\\
	108 & 6.91$\pm$0.21  & - & 3.8$\pm$2.2 & 6.99$\pm$0.19 & - & 6.76$\pm$0.13 & 7.39$\pm$0.09\\
	130 & 6.49$\pm$1.30  & - & - & 7.33$\pm$0.20 & - & 7.25$\pm$0.18 & - \\
	132 & 8.87$\pm$0.19  &- & - & 7.27$\pm$0.22 & - & 7.17$\pm$0.14 & 7.76$\pm$0.15\\
	139 & 6.57$\pm$1.20 &- & - & 7.42$\pm$0.26 & 10 & 7.12$\pm$0.16 & 7.53$\pm$0.08\\
\hline\hline 
\end{tabular}
\end{table*}

\section{Observation and Data Reduction}

We observed NGC\,6940 and To\,5 stars with the Main Stellar Spectrograph (MSS)\footnote{www.sao.ru/hq/lizm/mss/en/index.html} of the 6 meter telescope BTA  (Big Telescope Alt-azimuth) at the Special Astrophysical Observatory in Nizhny Arkhyz, Russia. This instrument is installed in the Nasmyth-2 focus of the telescope. It is essentially a long-slit spectrograph, which in its standard mode is equipped with a differential circular polarization analyser, and  combined with a double image slicer designed by Chountonov (\citeyear{2004choun}) and a rotating l/4 phase plate, to study mainly stellar magnetic fields.\
For the purpose of our observations, the instrument was set-up in its basic long slit mode. In this set-up, the MSS instrument allows to cover the wavelength range 5840$-$6690\AA~ with a resolution $\lambda / \Delta \lambda$ of $\sim$ 13000 .\\
Observations were carried out on the nights of September 7 and 8, 2017. The nights were clear and stable, with typical seeing in the range 1.5$-$3.0 arcsecs.
Beside the spectra of scientific targets, we collected bias and dome flat frames (two sets of 10 images obtained at the beginning and at the end of the night) and a set of ThAr spectra, for wavelength calibration purposes.\\

\noindent
Spectra were reduced in a standard  way, which consists of the following steps: 1) construction of the master bias and subtraction of it from the scientific and calibration frames, 2) correction of data for stray light, 3) searching for the location of individual slices in 2D images, 4) extraction of 1D spectra, 5) continuum normalisation of the spectra, 6) correction of wavelengths for the Earth's motion. All these procedures were carried out within ESO-MIDAS package  and its Zeeman extension. Spectra were normalized by the continuum level using the task  \textit{continuum} within the \texttt{IRAF} package\footnote{\texttt{IRAF} is distributed by the National
Optical Astronomy Observatory, which is operated by the Association of
Universities for Research in Astronomy, Inc., under cooperative agreement
with the National Science Foundation}. Wavelength solution was defined using the arc spectrum obtained closest in time to the scientific exposures.\\
An excerpt of the stars spectra is shown in Fig.\ref{Fig2}.\\

\section{Radial velocities}

Radial velocities were measured using the task \texttt{fxcor} of \texttt{IRAF}, which performs a Fourier cross-correlation between the observed spectra and a synthetic one. We chose a synthetic spectrum of a typical giant star with solar metallicity and more or less the same resolution (T$\mathrm{_{eff}}$=4750 K, log(g)=2.25 dex, $\xi$=2 km/s and R$\sim$12000) (Munari et al. \citeyear{2005mun}). The typical errors given by the task is of the order of 1.0 km/s.\\
In Tab.\ref{Tab3} and \ref{Tab4} the velocity measurements are reported for NGC\,6940 and To\,5, respectively. In the tables we reported also the values found in literature for a comparison. For all the other stars, our measurements are all in good agreement with the previous studies.\\
In the case of To\,5 stars, there are no previous sprectroscopic studies to compare our measurements with. However, two of them (\#5274 and \#8099) seem to be field stars. According to Zdanavi$\mathrm{\check{c}}$ius et al.(\citeyear{2011zda}), in the cluster region there are seven possible red giant candidates: five of them are in common with the ones we observed, except for \#5274 and \#8099. So we decided to exclude them from the chemical abundances analysis.\\
The mean radial velocities for the two clusters are V$\mathrm{_{r,NGC6940}}$=8.0$\pm$0.2 km/s, in good agreement with the values found in literature, and V$\mathrm{_{r,To5}}$=-22.8$\pm$0.4 km/s.\\

\begin{table}
\tabcolsep 0.65truecm
\caption{Radial velocity measurements for the program stars in To\,5.}
\label{Tab4}
\centering
\begin{tabular}{lc}
\hline\hline
ID &	$V_r$ \\
\hline
& \footnotesize{(km/s)}\\
\hline
0006 & -22.9$\pm$0.7\\
	5274 & -16.0$\pm$0.7\\
	5521 & -23.3$\pm$1.3\\
	7701 & -23.3$\pm$0.6\\
	7834 & -20.9$\pm$1.1\\
	8080 & -21.2$\pm$1.8\\
	8099 & -18.7$\pm$2.5\\
\hline\hline
\end{tabular}
\end{table}

\begin{table*}
\tabcolsep 0.4truecm
\caption{Atmospheric parameters for member stars in the two clusters. The first five columns report the values estimated from photometric data, while the last three are the values derived from our spectroscopic investigation.}
\label{Tab5}
\centering
\begin{tabular}{lcccccccc}
\hline\hline
ID   & T(B-V) & T(V-K) & $\mathrm{T_{eff,phot}}$ & $\mathrm{log(g)_{phot}}$ &  $\xi\mathrm{_{phot}}$ & $\mathrm{T_{eff,spec}}$& $\mathrm{log(g)_{spec}}$ & $\xi\mathrm{_{spec}}$\\

 & \footnotesize{(K)} & \footnotesize{(K)} & \footnotesize{(K)}  & \footnotesize{(dex)} & \footnotesize{(km/s)} & \footnotesize{(K)}  & \footnotesize{(dex)} & \footnotesize{(km/s)}\\
\hline
\textbf{NGC\,6940} &\\
28	& 4996 & 5068 &	5032 & 2.81 & 1.32 & 4980 & 2.91 & 1.32\\ 
60	& 5017 & 5096 & 5057 & 2.83 & 1.31 & 5057 & 2.85 & 1.31\\ 
67  & 4975 & 5034 & 5005 & 2.65	& 1.37 & 5005 & 2.55 & 1.37\\ 
108	& 5171 & 5225 & 5198 & 2.75	& 1.33 & 5100 & 2.70 & 1.33\\
130	& 5104 & 5175 & 5140 & 2.80	& 1.32 & 5140 & 2.80 & 1.32\\
132	& 5017 & 5127 & 5072 & 2.59	& 1.39 & 5000 & 2.70 & 1.39\\
139	& 5082 & 5153 & 5118 & 2.77	& 1.39 & 4980 & 2.95 & 1.39\\
\hline
\textbf{To\,5} &\\
0006 & 4979 & 5068 & 5024 & 2.36 & 1.46 & 5024 & 3.00 & 1.46\\
5521 & 5095 & 5398 & 5247 & 2.43 & 1.44 & 5150 & 3.40 & 1.44\\
7701 & 4558 & 4863 & 4711 & 2.05 & 1.56 & 4710 & 2.10 & 1.56\\
7834 & 5266 & 5363 & 5315 & 2.72 & 1.35 & 5215 & 2.80 & 1.35\\
8080 & 5058 & 5380 & 5219 & 2.61 & 1.38 & 5220 & 3.40 & 1.38\\
\hline\hline
\end{tabular}
\end{table*}

\section{Atmospheric parameters}

Atmospheric parameters were obtained as follows.\\
First, T$_{\mathrm{eff,phot}}$ estimates were derived from the B-V and V-K colours using the relations by Alonso et al. (\citeyear{1999alo}). In the chemical abundances analysis we used the mean temperature between T(B-V) and T(V-K) as input.\\
Surface gravities (log(g)) were obtained from the canonical equation:

\begin{center}
\begin{equation*}
\begin{split}
log(g)=& log(g)_{\odot}+log\left(\frac{m_{\star}}{m_{\odot}}\right)+4\cdot log\left(\frac{T_{\rm eff}}{T_{\rm eff,\odot}}\right)+\\
&+0.4\cdot(M_V+BC_V-M_{BC,\odot})
\end{split}
\end{equation*}
\end{center}

where T$_{\rm eff,\odot}$=5777\,K, log(g)$_{\odot}$=4.44\,dex, M$_{BC,\odot}$=4.74.\\
The bolometric correction (BC$_V$) for each star was derived by adopting the relations 
from Alonso et al. (\citeyear{1999alo}).
We adopted initial values for reddening E(B-V) and apparent distance modulus ${(m-M)_{\rm V}}$, from WEBDA, and inferred stellar masses (m$_{\star}$) from the Padova suite of isochrones (Marigo et al. \citeyear{2017mar}). The values used for each cluster are reported in Tab. \ref{Tab2}.\\
Micro-turbulent velocity ($\xi$) was obtained from the
relation by Gratton et al. (\citeyear{1996grat}):\\

\begin{center}
{\rm $\xi$= 2.22 - 0.322$\cdot$log(g)}
\end{center}

\noindent
The input metallicity needed to obtain T$_{\rm {eff,phot}}$ from colors and for the
isochrone fitting was assumed to be solar, which was later confirmed by the spectroscopic analysis (see below). The derived atmospheric parameters are listed in Tab.\ref{Tab5}.\\

\section{Elemental abundances}

\begin{table*}
\tabcolsep 0.65truecm
\caption{Derived abundances ratios for single stars in NGC\,6940 and To\,5. $\sigma_1$ and $\sigma_2$ are respectively the errors estimated from the EW measurements and from the dependences of the atmospheric parameters.}
\label{other}
\centering
\begin{tabular}{lcccccccc}
\hline\hline
ID  & [Si/Fe]$\pm\,\sigma_1$ $\pm\,\sigma_2$ & [Ca/Fe]$\pm\,\sigma_1$ $\pm\,\sigma_2$ & [Ti/Fe]$\pm\,\sigma_1$ $\pm\,\sigma_2$ & [Ni/Fe]$\pm\,\sigma_1$ $\pm\,\sigma_2$\\
\hline
\textbf{NGC\,6940} & \\
28	& 0.18$\pm$0.08 $\pm$0.17 & 0.09$\pm$0.35 $\pm$0.04 & -0.13$\pm$0.24 $\pm$0.11 & -0.01$\pm$0.17 $\pm$0.05\\
60	& -0.02$\pm$0.05	$\pm$0.15&	-0.07$\pm$0.34$\pm$0.03&	-0.11$\pm$0.19$\pm$0.10&	-0.2$\pm$0.18$\pm$0.14\\
67	& 0.07$\pm$0.04$\pm$0.18&	 0.19$\pm$0.09$\pm$0.12&	 0.03$\pm$0.13$\pm$0.09&	-0.1$\pm$0.09$\pm$0.05\\
108	& 0.14$\pm$0.07$\pm$0.16	 &0.07$\pm$0.12$\pm$0.01&	 0.04$\pm$0.14$\pm$0.10&	 0.14$\pm$0.11$\pm$0.05\\
130	 &0.13$\pm$0.06$\pm$0.14	 &0.09$\pm$0.1$\pm$0.02&	 0.04$\pm$0.13$\pm$0.09&	 0.23$\pm$0.09$\pm$0.03\\
132	 &0.2$\pm$0.05$\pm$0.16&	 0.08$\pm$0.04$\pm$0.02	& 0.06$\pm$0.07$\pm$0.11&	 0.04$\pm$0.06$\pm$0.04\\
139	 &0.12$\pm$0.14$\pm$0.16&	-0.15$\pm$0.18$\pm$0.04	&-0.22$\pm$0.1$\pm$0.14	& -0.03$\pm$0.07$\pm$0.05\\
\hline
\textbf{To\,5}&\\
0006 & 0.13 $\pm$0.13 $\pm$0.15	& 0.04 $\pm$0.03 $\pm$0.05	& -0.02 $\pm$0.09 $\pm$0.12 &	0.05 $\pm$0.14 $\pm$0.04\\
5521 & 0.10 $\pm$0.03 $\pm$0.15 & 0.16 $\pm$0.17 $\pm$0.04 & 0.13 $\pm$0.05 $\pm$0.11 &  -0.03 $\pm$0.05 $\pm$0.03\\
7701 & 0.04 $\pm$0.07 $\pm$0.18 & -0.07 $\pm$0.02 $\pm$0.08 & -0.09 $\pm$0.12 $\pm$0.11 & 0.08 $\pm$0.12 $\pm$0.09\\
7834 & -0.17 $\pm$0.04 $\pm$0.19 & 0.12 $\pm$0.11 $\pm$0.03 & 0.10 $\pm$0.16 $\pm$0.06 & 0.01 $\pm$0.19 $\pm$0.05\\
8080 & 0.00 $\pm$0.10 $\pm$0.15 & -0.09 $\pm$0.12 $\pm$0.07 & 0.04 $\pm$0.18 $\pm$0.08 & -0.04 $\pm$0.04 $\pm$0.03\\

\hline\hline 
\end{tabular}
\end{table*}

\begin{figure*}
\centering
\begin{minipage}[c]{19cm}
\centering
\includegraphics[scale=0.24]{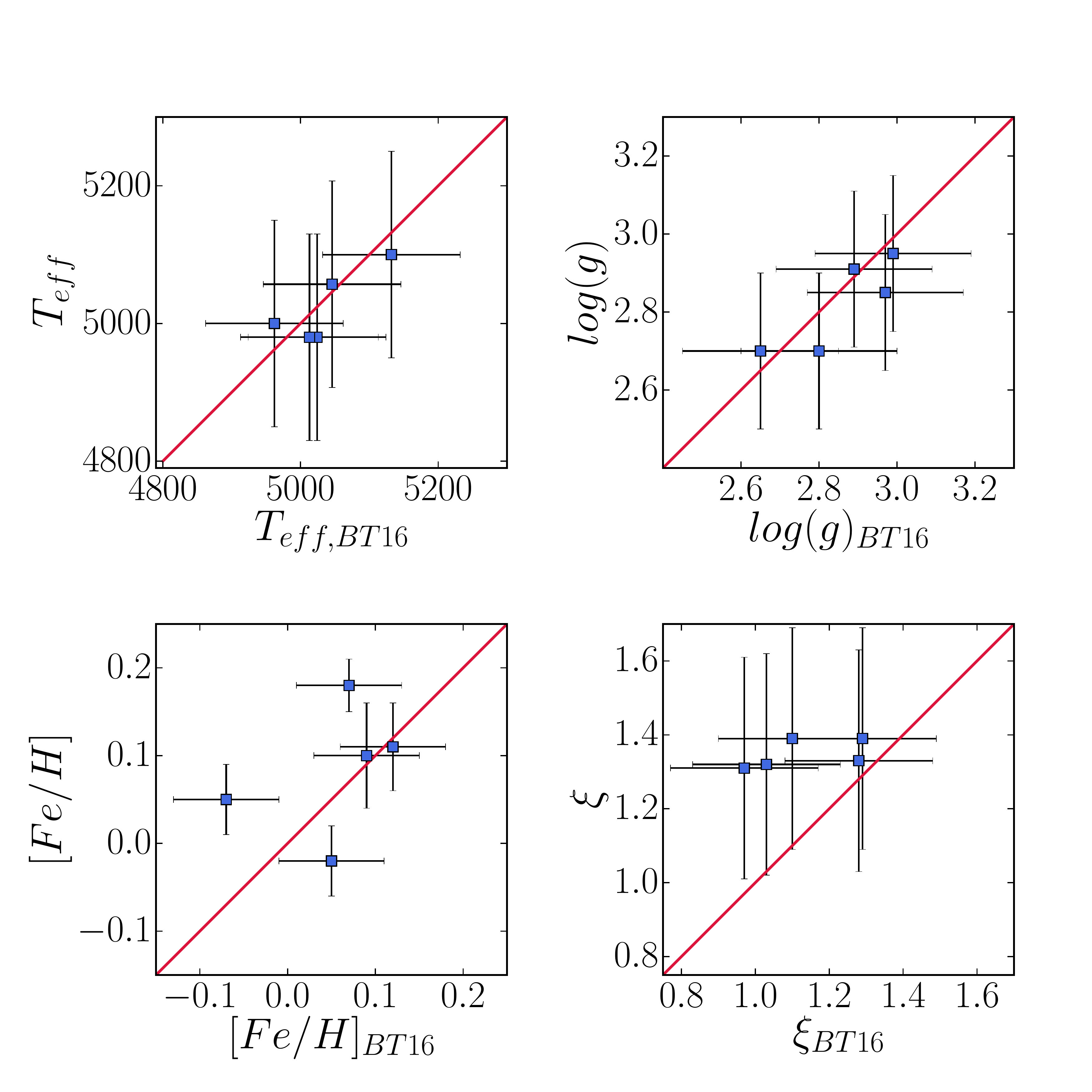}\hfil
\includegraphics[scale=0.24]{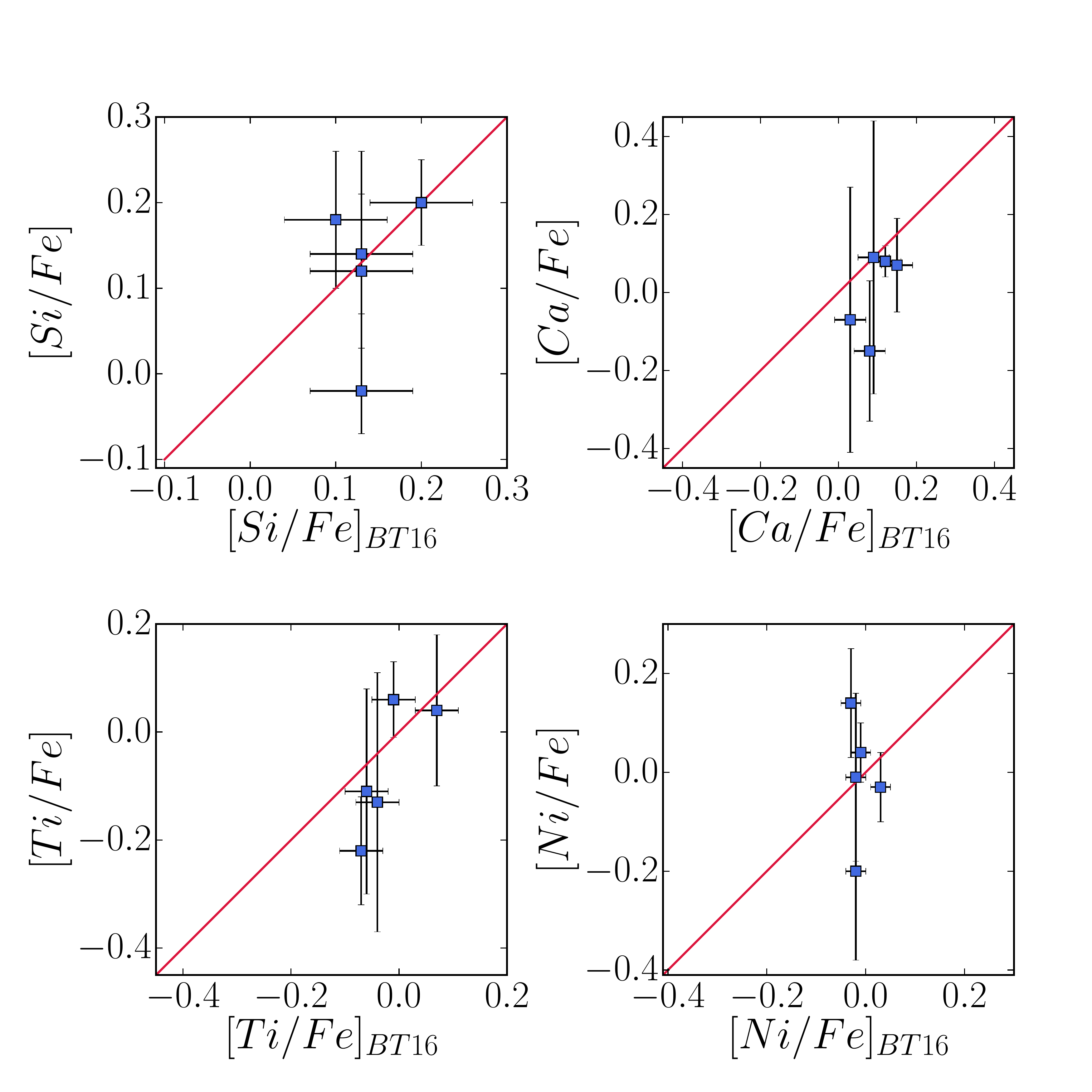}
\caption{Comparison between the atmospheric parameters (y-axis) and the values found by B$\mathrm{\ddot{o}}$cek Topcu et al. (\citeyear{2016boc}) (BT16, in x-axis). On the right panel we show the comparison between the other elemental abundances derived in this work (y-axis) and the same authors. A part for the micro-turbolence velocities, for which we considered valid the photometric values, the measurements are in agreement.}
\label{Fig3}
\end{minipage}
\end{figure*}

\begin{table}
\tabcolsep 0.25truecm
\caption{Derived metallicities for single stars in NGC\,6940 and To\,5. $\sigma_1$ and $\sigma_2$ are respectively the errors estimated from the EW measurements and from the dependences of the atmospheric parameters.}
\label{metN}
\centering
\begin{tabular}{lc}
\hline\hline
ID   & [Fe/H] $\pm \sigma_1$ $\pm \sigma_2$ \\
\hline
\textbf{NGC\,6940}&\\
28 & -0.02 $\pm$0.04 $\pm$0.17\\
60	& 0.11 $\pm$0.05 $\pm$0.16\\
67	&  0.15 $\pm$0.03 $\pm$0.2\\
108	& 0.05 $\pm$0.04 $\pm$0.18\\
130	&  0.04 $\pm$0.04 $\pm$	0.16\\
132	& 0.18 $\pm$0.03 $\pm$0.18\\
139	&  0.10 $\pm$0.06 $\pm$	0.16\\
\hline
\textbf{To\,5}&\\
0006 & 0.11	$\pm$0.04 $\pm$0.16\\
5521 & -0.05 $\pm$0.04 $\pm$0.16\\
7701 & 0.21	$\pm$0.05 $\pm$0.2\\
7834 & -0.04 $\pm$0.06 $\pm$0.2\\
8080 & 0.06	$\pm$0.11 $\pm$0.18\\
\hline\hline 
\end{tabular}
\end{table}

To derive the spectroscopic atmospheric parameters and chemical abundances of the stars confirmed members, we used the local thermodynamic equilibrium (LTE) line analysis and synthetic spectrum code \texttt{MOOG} (Sneden \citeyear{sne73}, Soebeck et al \citeyear{sob11} \footnote{https://www.as.utexas.edu/~chris/moog.html}.  
We used 1D model atmospheres linearly interpolated from the ATLAS9 model atmosphere grid of Castelli \& Kurucz (\citeyear{2003cast}).\\
Si\,I, Ca\,I, Ti\,I, Fe\,I, Fe\,II and Ni\,I abundances were estimated using the equivalent width (EQW) method by running the \textit{abfind} driver of \texttt{MOOG}.  EQWs for a selected line list were measured manually using the task \texttt{splot} in \texttt{IRAF}. The line list comes from D'Orazi et al. \citeyear{2017dor}. We kept only lines which are isolated and unblended, fitting a gaussian profile and/or by direct integration, depending on the feature under consideration. We considered only lines with EQWs$<$150\,m\AA. \\

\noindent
The final temperatures were derived by removing the trend between the iron abundances (log(Fe\,I)) of the single lines and the excitational potential ($\chi$). We changed the temperature until the slope of the trend was compatible with a flat trend within 1$\sigma$. In most cases, the input values were close to the final ones, so no changes were applied to the temperatures. As for the final value of the surface gravity, we checked the difference between log(Fe\,I) and log(Fe\,II) and changed it until it reached a value below 0.05 dex. As for micro-turbolence velocities, we did not change the input value since the numbers of Fe\,I lines were not sufficient to calculate a meaningful linear trend between the abundances and the reduced equivalent widths (REWs). The final metallicity  were computed with respect to the solar value and reported in Tab.\ref{metN}.\\
We calculated also the abundances of all the others elements and then the abundance ratios with respect to Fe\,I. The final values are reported in Tab.\ref{other}.\\

\noindent

In order to estimate the uncertainties of the metallicities and abundance ratios, we considered mainly errors due to EQWs measurements and to internal uncertainties of the stellar parameters, defined as $\sigma_1$ and $\sigma_2$ in Tab.\ref{metN}.\\
The first source of error is well represented by the standard deviation from the mean abundances considering all the lines divided by the root square of the number of lines. The $\sigma_1$ values derived for [X/Fe] (reported in Tab.\ref{other}) were calculated by quadratically adding the $\sigma_1$ value of [Fe/H] and the one of [X/H].\\ 
The $\sigma_2$ values were estimated by varying T$_{eff}$ by 150\,K, log(g) by 0.2\,dex and $\xi$ by 0.3\,dex. For each new model, we calculated the difference between the new [X/Fe] and that obtained with the final parameters. Then we added all the contributions quadratically.\\

\begin{figure}
\centering
\includegraphics[width=.45\textwidth]{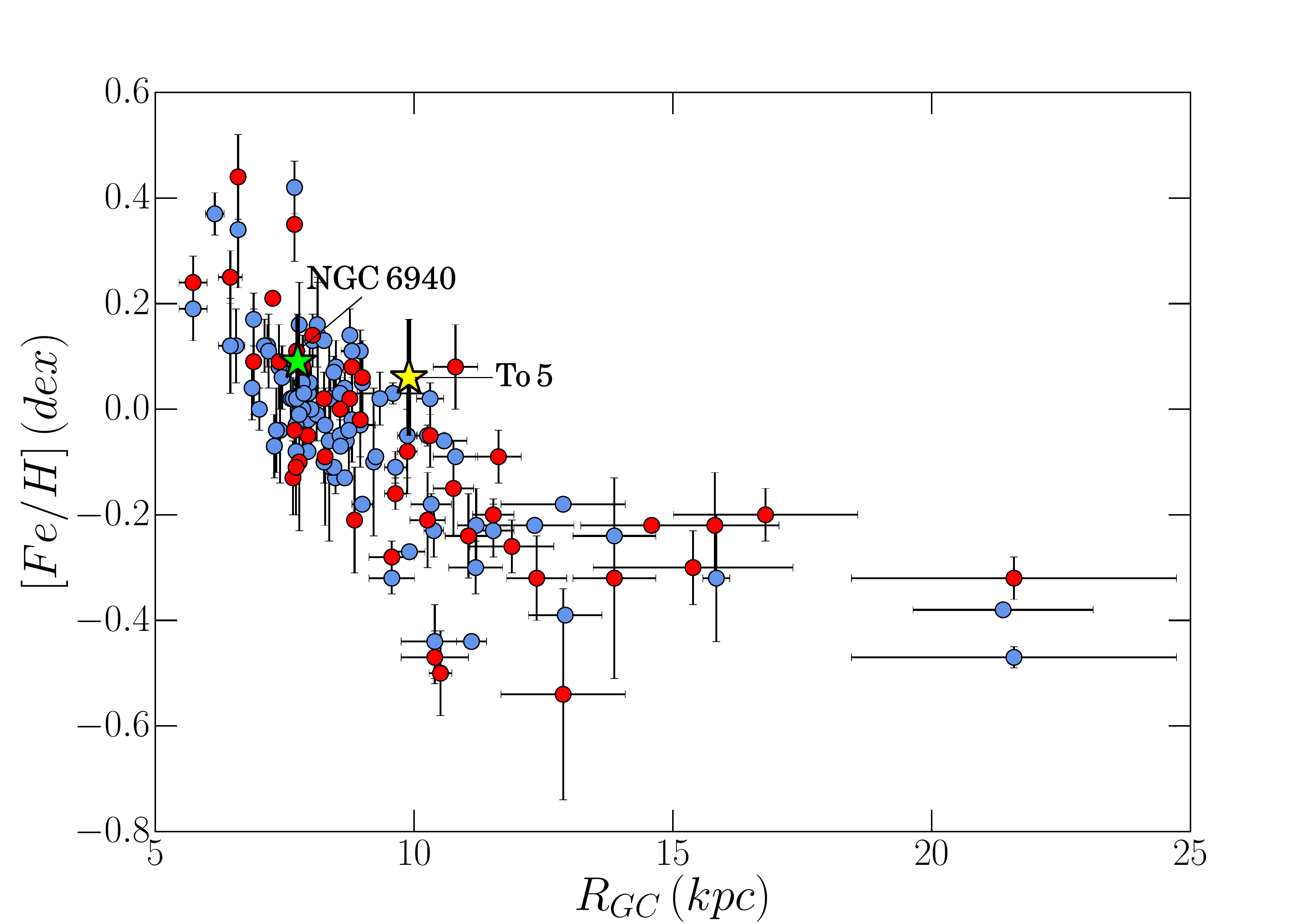}
\caption{Galactic radial metallicity gradient from Netopil et al. (\citeyear{2016neto}), considering only open clusters with spectroscopic determination of the metallicity (red dots stand for low quality spectroscopic data, while blue ones for high quality spectroscopic data.). Starred symbols indicate the location of our two clusters.}
\label{Fig5}
\end{figure}

\section{Results of the chemical analysis}

The final values of the atmospheric parameters and the individual abundances are reported in Tabs. \ref{Tab5} and \ref{other}. In order to evaluate the validity of our method, we checked the correlation between the derived metallicities and the atmospheric parameters. Overall we could not find any evident trend.\\
In the case of NGC\,6940, we could compare our results with the values derived by B$\mathrm{\ddot{o}}$cek Topcu et al.(\citeyear{2016boc}) for five stars in common. The mean values of the cluster metallicities and abundances ratios are reported in Tab.\ref{Tab9}, in which we also reported the literature values for a comparison.\\

\begin{table*}
\tabcolsep 0.3truecm
\caption{Comparison of mean metallicity and abundances ratios derived in this work with the literature.}
\label{Tab9}
\centering
\begin{tabular}{lcccccccc}
\hline\hline
ID	&$<[Fe/H]>$&	$<[Si/Fe]>$&	$<[Ca/Fe]>$	&$<[Ti/Fe]>$&	$<[Ni/Fe]>$\\
\hline
\textbf{NGC\,6940} &\textbf{0.09$\pm$0.06}	& \textbf{0.11$\pm$0.07}&	\textbf{0.04$\pm$0.11}&	\textbf{-0.04$\pm$0.10}& \textbf{0.01$\pm$0.13}\\
Blanco-Cuaresma et al.(2015)& 0.04$\pm$0.09& 0.13$\pm$0.13 & 0.03$\pm$0.09 & -0.01$\pm$ 0.13& -0.05$\pm$0.10\\
Bocek-Topcu (2016)	&0.04$\pm$0.02&	0.12$\pm$0.06	&0.08$\pm$0.04	&-0.04$\pm$0.04&	-0.01$\pm$0.02\\
Reddy(2016)	&-0.06$\pm$0.03&	0.14$\pm$0.03&	0.00$\pm$0.03&	-0.04$\pm$0.02	&0.00$\pm$0.03\\
\hline
\textbf{To\,5}& \textbf{0.06$\pm$0.11} & \textbf{0.02$\pm$0.11}& \textbf{0.03$\pm$0.10}& \textbf{0.03$\pm$0.08}& \textbf{0.01$\pm$0.05}\\
\hline\hline 
\end{tabular}
\end{table*}

\subsection{NGC\,6940}

In Fig.\ref{Fig3} (left panel) we show the comparison between the atmospheric parameters derived in this work (y-axis) and the values found by B$\mathrm{\ddot{o}}$cek Topcu et al.(\citeyear{2016boc}) (BT16, in x-axis). The agreement is quite good: each measurement is compatible with the authors', except for the micro-turbulence velocities. We kept as valid the photometric estimates of these values, since we had too few lines to calculate the trend between the individual abundances and the reduced equivalent widths.\\
In the right panel of Fig.\ref{Fig3}, instead, we compared the abundance ratios. Some values we found are compatible with B$\mathrm{\ddot{o}}$cek Topcu et al.(\citeyear{2016boc}), but the dispersion of our measurements is larger. This can be explained with the lower resolution of our spectra and the limited spectral coverage. Moreover, most of the lines do not have a pure gaussian profile, so we measured the EQWs with other methods.\\
The mean value of metallicity and abundance ratios, as reported in Tab.\ref{Tab9}, are in very good agreement with the values found in literature. NGC\,6940 has a nearly solar metallicity of +0.09$\pm$0.06\,dex, consistent with the values expected for a cluster at that Galacto-centric distance, as Fig.\ref{Fig5} shows. he Galacto-centric distance has bee computed adopting $d_{\odot}$ =8.5 kpc, for homogeneity with previous studies and for the sake of comparison.\\

\subsection{To\,5}

Basing on previous estimates of the distance, this cluster appears to be quite interesting, since it would be located close to the critical Galacto-centric distance at which the radial abundance gradient for [Fe/H] exhibits a 
significant change of slope.
For this cluster we provided the very first spectroscopic estimate of the photospheric parameters for all the sample stars but  for  $\#$8080, for which we had to rely on the mean photometric estimates from the other sample stars.  
We derived for To\,5 a roughly solar metallicity of +0.06$\pm$0.11\,dex, which lies close to the upper envelope of the  thin disk  abundance gradient, but is still compatible with
the other clusters located in the same Galacto-centric distance range, as shown in Fig.\ref{Fig5}.

\section{Cluster basic parameters revisited}

\begin{figure*}
\centering
\includegraphics[scale=0.35]{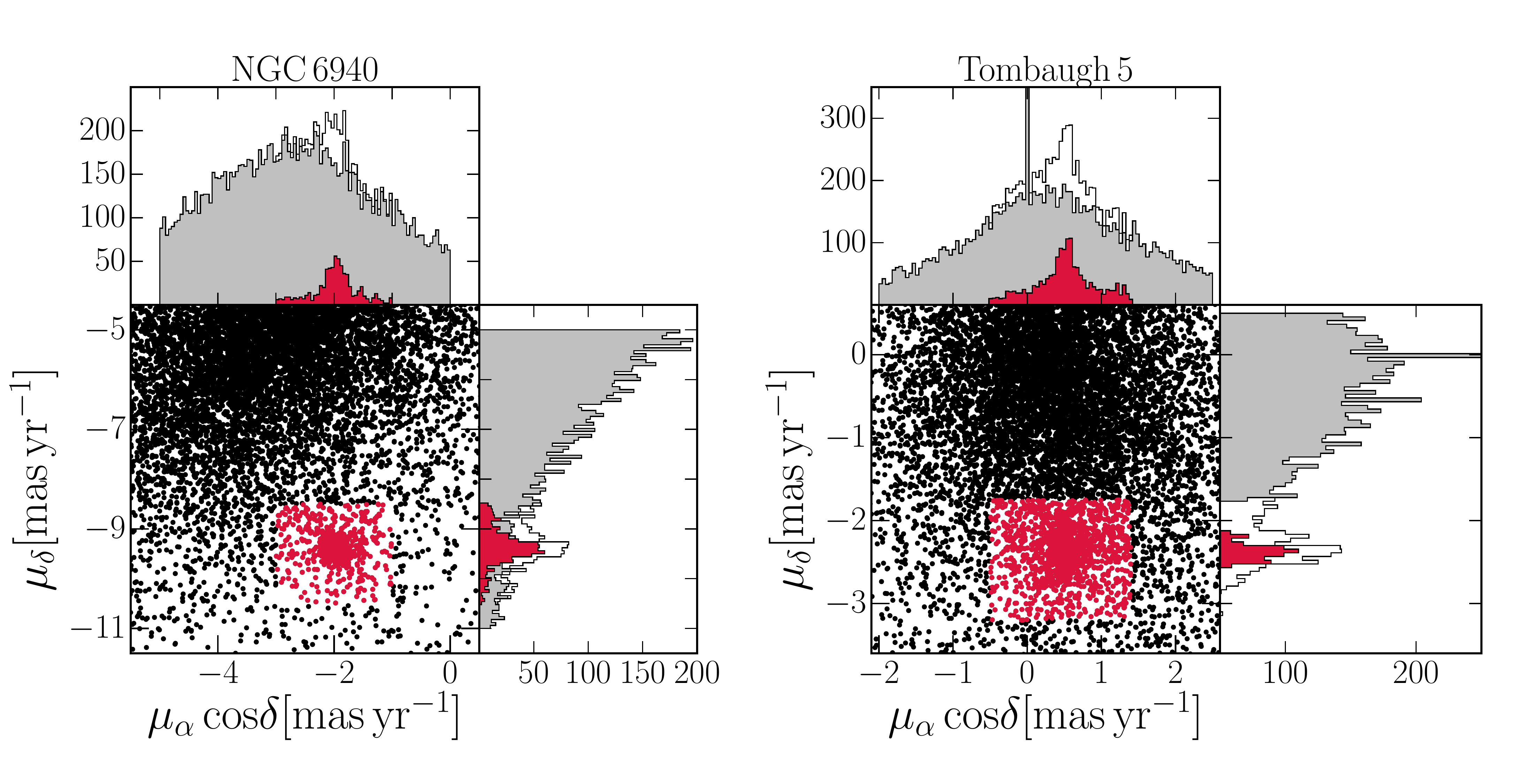}
\caption{Vector point diagrams of NGC\,6940 and To\,5: the black dots represent the data taken from Gaia DR2, while the red ones represent our conservative selection of cluster members. }
\label{Fig6}
\end{figure*}

\begin{table*}
\tabcolsep 0.3truecm
\caption{Proper motion components, parallaxes and distances derived from \textit{Gaia} DR2, and corresponding errors, for the stars in common with this study.}
\label{Tab10}
\centering
\begin{tabular}{lcccccccc}
\hline\hline
ID	&$\mu_{\alpha} cos\delta \pm \sigma_{\alpha}$&	$\mu_{\delta} \pm \sigma_{\delta}$&	$\varpi \pm \sigma_{\varpi}$	&$d \pm \sigma_d$\\
\hline
\textbf{NGC\,6940} \\
67 & -1.758$\pm$0.061 & -9.514$\pm$0.046& 1.026$\pm$0.038& 975$\pm$ 36 \\
108 & -1.534$\pm$ 0.046 & -9.235$\pm$ 0.041 & 0.883$\pm$ 0.033 & 1132$\pm$ 42\\
130 & -2.178$\pm$ 0.049 & -9.182$\pm$ 0.047 & 0.934$\pm$ 0.035 & 1071$\pm$ 4\\
132 & -2.015$\pm$ 0.051 & -9.315$\pm$ 0.047 & 1.005$\pm$ 0.039 & 995$\pm$ 38\\
139 & -2.046$\pm$ 0.045 & -9.523$\pm$ 0.046 & 0.975$\pm$ 0.031 & 1026$\pm$ 33\\
\hline
\textbf{To\,5} \\
0006 & -7.478$\pm$0.862& 1.528$\pm$0.748& -3.061$\pm$0.497& 327$\pm$55\\
5521 &0.270$\pm$0.046& 4.233$\pm$0.042& 0.999$\pm$0.031& 1000$\pm$31 \\
7701 & 0.383$\pm$ 0.065 & -2.297$\pm$ 0.064 & 0.522$\pm$ 0.039 & 1917$\pm$ 145\\
7834 & -3.368$\pm$0.048& -0.923$\pm$0.048&1.007$\pm$0.031&993$\pm$31\\
8080 & 23.889$\pm$1.727 & -20.752$\pm$1.632& -1.178$\pm$1.155&849$\pm$130\\
\hline\hline 
\end{tabular}
\end{table*}

\begin{figure}
\includegraphics[scale=0.3]{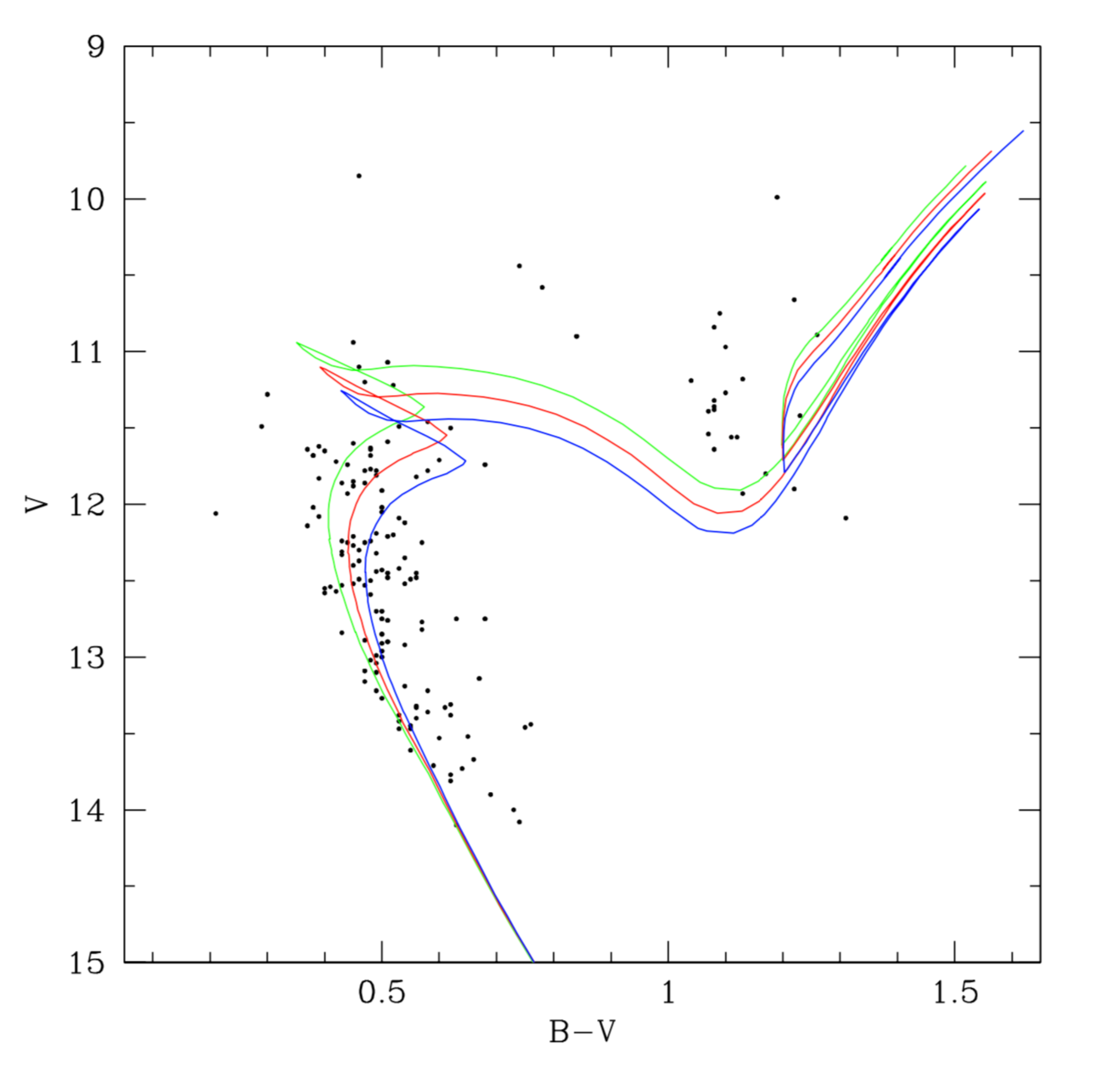}
\caption{Color magnitude diagrams of NGC\,6940 from the BV study of Larsson-Leander (\citeyear{1960lar}). Isochrones are super-imposed for a distance of 1.0 kpc and a reddening of 0.21 mag, and ages of 0.9 (green), 1.0 (red), and 1.1 (green) Gyr. See text for additional details.}
\label{Fig7}
\end{figure}

\begin{figure}
\includegraphics[width=\columnwidth]{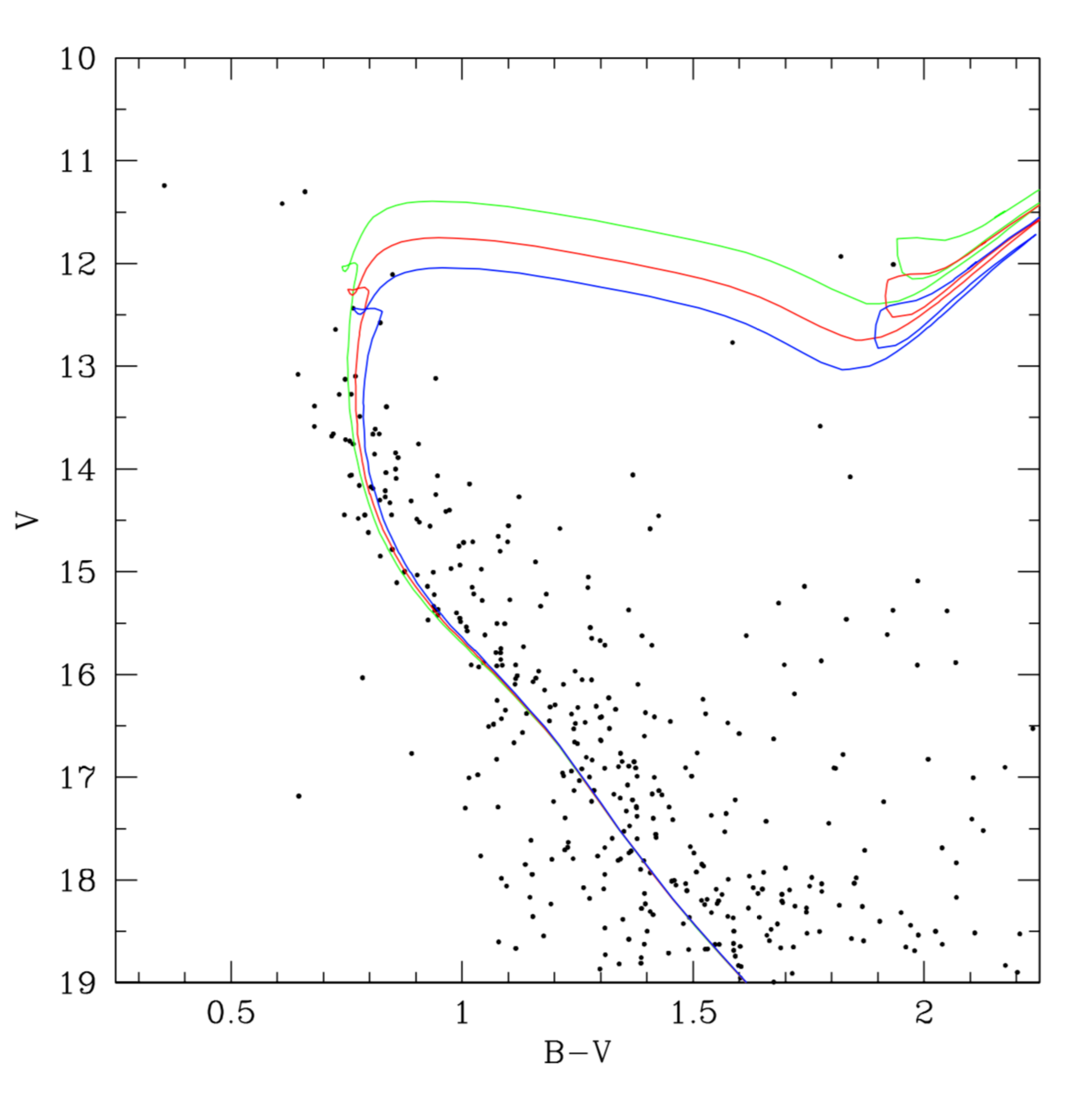}
\caption{Color magnitude diagrams of To\,5 from the BV study of Maciejewski \& Niedzielski (\citeyear{2007mac})).  Isochrones are super-imposed for a distance of 1.7 kpc, a reddening of 0.80 mag,  and ages of 200 (green) , 250 (red), and 300 (blue) Myr. See text for additional details. }
\label{Fig8}
\end{figure}

In order to estimate the cluster basic parameters, we complement available photometric and spectroscopic data with the  recent \textit{Gaia} DR2 data. 
As in Fig. \ref{Fig6}, we plotted proper motion $\mu_{\alpha}cos(\delta)$ (mas/yr) vs $\mu_{\delta}$ (mas/yr)  and selected visually only the stars within the cluster radius. Then we calculated the distance from the parallax measured by \textit{Gaia}  only for  the proper-motion-selected stars.\\
 As one can appreciate from Fig. \ref{Fig6}, the distributions have a gaussian profile around the mean value. To get a distance estimate we selected all the stars within 3$\sigma$ from the mean to include the largest number of candidate  members.  We note,
however, that two NGC~6940 stars observed by us are not included  \textit{Gaia} DR2 (namely \#28 and \#60), while the majority of the stars in the red clump of To\,5 are missing, either because their distances are not within 3$\sigma$ from the mean (like \#0006) or because they have peculiar proper motion (like \#8080). All of them, however, are confirmed members according to radial velocities and metallicity.
The values of proper motion components, parallaxes, derived distances and the mean cluster values are reported in Tab.\ref{Tab10}.\\
A few recent studies have noted that the zero-point of Gaia DR2 parallaxes should probably be incremented by 0.05-0.08 mas (Stassun \& Torres. \citeyear{sta18} , Zinn et al. \citeyear{zin18} . We will investigate this possibility for To5, for which distance is not yet well-constrained.\\

\noindent
To provide an updated estimate of the clusters age,  we super-imposed isochrones for different ages on top the CMD built with optical BV photometry as in Fig.~1. The isochrone fitting is performed 
fixing the metallicity to the new value found in this work and  the reddening to the value found in literature (see Fig.\ref{Fig7} and \ref{Fig8} for NGC~6940 and To\,5, respectively). \\

\noindent
In the case of NGC\,6940 the fit
yields an age of 1.0$\pm$0.1\,Gyr (for a metallicity Z=0.0187) in nice agreement with B$\mathrm{\ddot{o}}$cek Topcu et al. (\citeyear{2016boc})) adopting the  \textit{Gaia} distance estimate of 1.03$\pm$0.17\,kpc , which is slightly higher than the values found in literature and reported in WEBDA.  This corresponds to an apparent distance modulus of (m-M)$_V$=10.72$\pm$0.36.\\

\noindent
The distance estimate of To\,5 reported in WEBDA database is confirmed by \textit{Gaia} DR2, from which we  derive an estimate of 1.7$\pm$0.1\,kpc.  If we apply an offset of 0.05 mas, the distance becomes 1.68, well within the error bars.

This was calculated using MS  proper-motion-selected stars brighter than 16.0  in the \textit{Gaia} passband $G_{bp}$, since for our sample of clump stars \textit{Gaia} provides extremely different proper motion components and parallaxes. We do not have a clear explanation for this, since these stars have the same radial velocity and metallicity. Since To\,5 field is severely contaminated by field stars, one possibility is that there are mismatches in \textit{Gaia} DR2, which will probably be solved with future releases of the catalog. Another possibility is that these 5 stars have similar properties by chance, and they do not have anything to do with the cluster. We would exclude this scenario since their position in the cluster CMD is compatible with the expected locations of red, evolved stars in their He-burning phase.  To better clarify this issue,
we run a few simulations with the Trilegal code (Girardi  \citeyear{leo16}), and found that  for a single stellar populations (SSP) of the age nad estimated mass ($\sim 350 M_{\odot}$) of To5, the expected number of evolved stars is 4$\pm$3, well in agreement with available photometric data (see Fig~1).\\

\noindent
Using this distance, our new estimate of metallicity (which corresponds to Z=0.0175), and the reddening estimate provided by Majaess et al. (\citeyear{2008maj}) we derived a new age estimate using the isochrone fitting method, as shown in Fig. \ref{Fig8},
where a CMD based on Maciejewski \& Niedzielski (\citeyear{2007mac}) has been plotted.
We find that an age of 250$\pm$50 Myr nicely fits the overall stars' distribution, thus supporting earlier investigations.
The Gaia DR2 distance of 1.7 kpc places To\,5 in the close vicinity of the Perseus arm, as defined by the position of star forming regions
in this Galactic sector derived from maser distances as in Reid et al. (\citeyear{reid09}).

\section{Summary and Conclusions}

In this work we presented a spectroscopic study of two open clusters: NGC\,6940 and To\,5. While the former was studied before, for the latter these are the first spectroscopic measurements to-date. Our findings can be summarised as follows:

\begin{itemize}
\item We observed seven candidate red clump stars in the region of the two clusters. All the observed stars in NGC\,6940 turn out to be cluster members. The cluster mean value is V$\mathrm{_{NGC\,6940}}$=8.0$\pm$0.2 km/s. In the case of To\,5, we could confirm the membership of five of the observed stars and the mean radial velocity is V$\mathrm{_{r,To5}}$=-22.8$\pm$0.4 km/s.
 
\item  We then derived abundances for the following elements: FeI, SiI, CaI, TiI and NiI. Overall, the two clusters exhibit solar composition, with mean metallicity  [Fe/H]$\mathrm{_{NGC\,6940}}$=+0.09$\pm$0.06\,dex, and [Fe/H]$\mathrm{_{To\,5}}$=+0.06$\pm$0.11\,dex. The estimate we found for NGC\,6940 is in good agreement with the most recent literature. On the other hand, the metallicity measured for To\,5 confirms the large spread (almost 0.5 dex) in [Fe/H] at about10 kpc from the Galactic centre. 

\item We used \textit{Gaia} DR2 to select the cluster members based on the analysis of proper motions and to calculate the cluster distances from  the corresponding parallaxes. We derived for NGC\,6940 a heliocentric distance of 1.0$\pm$0.2\,kpc. This is somewhat larger than the values found in literature, but provides a generally good fit of the star distribution in the CMD.  For To\,5 we found a distance of 1.7$\pm$0.1\,kpc, confirming the values reported in WEBDA, and thus ruling out previous suggestions of a  shorter distance of  $\sim$1.3 kpc.

\item With our new estimates of metallicities and an updated distances from \textit{Gaia} DR2, we revised the cluster fundamental parameters, in particular the age. Using the isochrone fitting technique we found an age of 1.0$\pm$0.1\,Gyr for NGC\,6940,  and 250$\pm$50\,Myr for To\,5.
\end{itemize}

\acknowledgments
G. Carraro thanks the SAO staff, in particular Gennady Valyavin, for the kind hospitality. The authors also thanks Ulisse Munari  for enlightening discussions.
We made extensive use of the WEBDA databased, maintained by E. Paunzen at Masaryk University in Brno. We acknowledge the useful input of the anonymous referee.

\vspace{5mm}
\facilities{CAO}

\software{IRAF ( Tody \citeyear{tody93,tody96}), MIDAS (Banse et al. \citeyear{ban83};  Grosbol \& Ponz \citeyear{gro90}) }


\allauthors

\listofchanges

\end{document}